\documentclass[onecolumn]{aa}
\usepackage{graphicx}
\usepackage{txfonts}
\begin{document}
   \headnote{Research Note}
   \title{Discovery of a Variable Star Population in NGC 2808}
	
   \author{T. Michael Corwin\inst{1}\fnmsep\thanks{Visiting Astronomer, 
               European Southern Observatory, La Silla, Chile}
          \and
           M. Catelan\inst{2}\fnmsep$^{*}$
          \and
           J. Borissova\inst{2}\fnmsep$^{*}$
          \and
           H. A. Smith\inst{3}
          } 

   \offprints{M. Catelan}
        
   \institute{Department of Physics, University of North Carolina at Charlotte, 
              Charlotte, NC 28223, USA\\
              e-mail: mcorwin@uncc.edu
              \and    
              Pontificia Universidad Cat\'olica de Chile, Departamento de 
              Astronom\'\i a y Astrof\'\i sica, \\
              Av. Vicu\~{n}a Mackenna 4860, 782-0436 Macul, Santiago, Chile\\ 
              e-mail: mcatelan, jborisso, ajd@astro.puc.cl
              \and
              Dept.\ of Physics and Astronomy, Michigan State University, 
              East Lansing, MI 48824, USA\\
              e-mail: smith@pa.msu.edu
              }

   \date{Received; accepted}

   \abstract{
We have applied the image subtraction method to images of the peculiar, 
bimodal-horizontal branch globular cluster NGC~2808, taken over a total 
of six nights over a range of five months. As a result, we have found, 
for the first time, a sizeable population of variable stars in the 
crowded inner regions of the cluster, thus raising the known RR Lyrae 
population in the cluster to a total of 18 stars. In addition, an 
eclipsing binary and two other variables with periods longer than 1~day 
were also found. Periods, positions and (differential) light curves are 
provided for all the detected variables. The Oosterhoff classification 
of NGC~2808, which has recently been associated with a previously 
unknown dwarf galaxy in Canis Major, is briefly discussed. 
   }

   \maketitle
%

\section{Introduction}

Adequate interpretation of the color-magnitude diagrams of globular clusters 
(GCs) in terms of age, metallicity and other evolutionary parameters is 
one of the most important goals of stellar astrophysics, constraining both
the formation history of the Galaxy and the age of the Universe. Yet, there 
are still several open problems in the interpretation of these diagrams, one 
of the most outstanding being the so-called ``second-parameter problem.'' 

The prominent sequence that is seen as a ``horizontal branch'' (HB) in 
the visual color-magnitude diagrams of Galactic GCs generally 
changes from being predominantly comprised of stars to the red of the RR Lyrae 
(RRL) instability strip at high metallicities to having mostly blue HB stars 
at the low end of the [Fe/H] scale. However, Sandage \& Wildey (1967) and 
van den Bergh (1967) noted that, at a given metallicity, one may find very 
different HB types. Perhaps the best known examples are the second-parameter 
pairs NGC~288/NGC~362 and M13/M3, comprised of GCs which have very closely 
the same metallicity, and yet dramatically different HB morphologies.

Scenarios for the formation of the Galaxy whereby it may have grown 
through the accretion of protogalactic fragments have been based, to a 
large degree, on the interpretation that age is the second parameter of 
HB morphology (e.g., Searle \& Zinn 1978; van den Bergh 1993; Zinn 1993). 
Indeed, there is now  
some strong evidence that age may play an important part in explaining the  
syndrome, at least in the paradygmatic case of NGC~288/NGC~362 (Catelan et 
al. 2001 and references therein). On the other hand, the case of M13/M3 
suggests that age is unlikely to be the sole explanation (Stetson 1998, 
2000). 

Indeed, several additional parameters have not been ruled out and may 
also play a role, including mass loss (Catelan 2000), primordial abundance 
variations (D'Antona et al. 2002), stellar rotation 
(Norris 1983), deep mixing on the red giant branch (RGB; Sweigart 1997), core 
density (Buonanno et al. 1997), and even planetary systems (Soker \& Harpaz 
2000). In this sense, 
it has also been argued that GCs presenting the second parameter 
syndrome {\em internally}~--~that is, those with {\em bimodal} HBs~--~could 
provide a key to understanding the phenomenon (e.g., Rood et al. 1993; 
Catelan et al. 1998a, 1998b). Conversely, a fully satisfactory solution to the 
problem will not logically have been reached until bimodal-HB GCs have been 
properly accounted for. 

RR Lyrae (RRL) stars may be able to assist us in our quest for an answer. 
In particular, some models for the origin of bimodal HBs require that 
blue HB and RRL stars be unusually bright, whereas others do not. 
On the other hand, the 
period-mean density relation tells us that the RRL pulsation period, at a 
given $T_{\rm eff}$ (or amplitude), is strongly sensitive to the star's 
luminosity. Therefore, if RRL variables in bimodal-HB GCs follow similar 
period-temperature relations as do the RRL in GCs that do not show the 
second-parameter effect, 
without significant {\em period shifts} that may be attributed to luminosity 
effects, their HB bimodality is likely to be caused by mass loss variations 
among RGB stars, and/or by internal age variations within the cluster. This
is due to the fact that both these 
parameters affect only the mass with which a star will arrive on the 
zero-age HB (ZAHB), differences in which may cause a star to ``slide'' 
horizontally along the $T_{\rm eff}$ axis but not shift vertically along 
the $\log L$ axis. Conversely, period shifts with respect to the RRL 
in ``well-behaved'' GCs should be indicative of second parameters that 
do directly affect HB luminosity (Catelan et al. 1998b).

Of all the bimodal-HB Galactic GCs, NGC~2808 (C0911-646) 
stands out as the best known, 
and most dramatic, example. Being one of the most luminous GCs in our 
galaxy, with $M_V = -9.39$~mag (Harris 1996), it was early noted to have 
very prominent blue and red HB components, with very few if any stars in 
between (Harris 1974, 1975, 1978). Searches for RRL stars in the cluster 
have indeed confirmed that the cluster is largely devoid of these variables, 
photographic work culminating with the paper by Clement \& Hazen (1989). 

However, appropriate searches utilizing CCD detectors have apparently never 
been conducted in this cluster. To date, the only CCD study tackling the 
cluster RRL to appear in the literature seems to have been 
the report by Byun \& Lee (1991), 
utilizing images obtained with the CTIO 0.9m telescope, which however does 
not seem to be based on time-series photometry. No actual RRL candidates 
were reported in that study. Accordingly, the Clement et al. (2001) 
catalogue currently lists a mere 5 variables in the cluster, only two of 
which are actually RRL stars~--~the remaining being a BL Herculis star 
and two semi-regular variables. 

Recent success at detecting sizeable numbers of RRL variables in crowded 
GC fields (e.g., Kaluzny et al. 2001; Borissova et al. 2001; Kopacki 2001; 
Corwin et al. 2003), based on high-quality CCD 
images and image subtraction techniques (ISIS: Alard 2000; Alard \& Lupton 
1998), has motivated us to revisit the variable star population in NGC~2808. 
Additional motivation is provided by the suggested association of NGC~2808 
with the recently discovered Canis Major dwarf galaxy (Martin et al. 2004), 
given the constraints that RRL stars may pose on the formation history of 
the Galaxy (Catelan 2004; Kinman et al. 2004). Accordingly, the present 
paper reports on the first detection of a sizeable population of variable 
stars in the cluster, mainly comprised of RRL stars.

\begin{table*}
 \caption{New variable stars detected in NGC~2808}
 \label{tab1}
 \begin{tabular}{lccllllll}
 Name & $x$ & $y$ & RA & DEC (J2000) & Period (d) & Type & Comment \\ 
 \hline 
V1  & +107    &  $-35$   & 9:12:20 &    $-64$:52:21.3 &           & SR   \\
V6  & +39     &  $-66$   & 9:12:30 &    $-64$:56:39.0 &   0.53897 & RRab \\
V10 & $-42$   &  $-98$   & 9:11:57 &    $-64$:53:24.0 &   1.76528 & Cepheid \\
V12 & $-45$   &  94      & 9:11:56 &    $-64$:50:10.2 &   0.30578 & RRc \\ 
V13 &   124.32 &  323.04 & 9:12:24 &	$-64$:57:11.4 &	0.21	& RRc  & RRe? W~UMa? \\ 
V14 & 79.48  &  $-50.60$ & 9:12:13 &	$-64$:51:04.3 &	0.60	& RRab \\
V15 & 60.92  &  $-39.60$ & 9:12:11 &	$-64$:51:15.1 &	0.61	& RRab \\
V16 &   46.32  &  56.64  & 9:12:09 &	$-64$:52:51.9 &	0.59	& RRab \\
V17 &   35.76  &  0.88   & 9:12:07 &	$-64$:51:57.4 &	0.38	& RRc  \\
V18 &   31.92  &  21.08  & 9:12:07 &	$-64$:52:17.2 &	0.58	& RRab & not phased in $V$ \\
V19 &   30.00  &  14.36  & 9:12:06 &	$-64$:52:10.5 &	0.51	& RRab \\
V20 & 27.48  &  $-25.96$ & 9:12:06 &	$-64$:51:30.9 &	0.29	& RRc  \\
V21 & 17.24  &  $-31.64$ & 9:12:04 &	$-64$:51:26.2 &	0.60	& RRab \\
V22 & 15.52  &  $-1.68$  & 9:12:04 &	$-64$:51:56.7 &	0.54	& RRab & large scatter in $V$ \\
V23 &   10.40  &  22.80  & 9:12:03 &	$-64$:52:19.6 &	0.27	& RRc  \\
V24 &   7.08   & 13.68   & 9:12:03 &	$-64$:52:11.4 &	0.27	& RRc  \\
V25 &$-16.52$ &  $-30.92$ & 9:11:59 &	$-64$:51:28.8 &	0.49	& RRab & Blazhko? \\ 
V26 &$-52.46$ &  $-30.12$ & 9:11:53 &	$-64$:51:30.8 &	0.37	& RRc  \\
V27 & $-53.64$ &   59.76  & 9:11:54 &	$-64$:53:00.0 &	0.57	& RRab \\
V28 &$-68.60$ & $-33.84$  & 9:11:51 &	$-64$:51:28.0 &	0.28	& RRc  \\
V29 & $-9.83$ & $-19.08$  & 9:12:00 &	$-64$:51:38.8 &	1.97	& Cepheid? \\
V30 & 39.09   & $-13.14$    & 9:12:08 &	$-65$:51:44.6 &	1.47	& EB  & $\beta$~Lyr? \\ 
 \hline 
 \end{tabular}
\end{table*}

\section{Analysis}
The CCD images used in this study were obtained with the Danish 1.54m 
telescope located at the European Southern Observatory at La Silla, 
Chile. The field was observed for a total of six nights, including two nights 
in October 2002 (14/15 and 15/16), two nights in December 2002 (11/12 and 12/13), 
and two nights in February 2003 (18/19 and 19/20). The seeing in October was 
not particularly good, generally around 1.5 to 2.0~arcsec, and images were 
generally elongated; whereas the December and Feburary seeing was much better, 
around 1.0~arcsec or less, and image elongation problems were not as severe. 
Unfortunately, the October $B$ data turned out to be severely affected by 
fringing. Consequently, our adopted image subtraction analysis (based on ISIS2.1; 
Alard 2000) of the October data either did not produce useful results or was 
unable to locate the variables found in the December and February data. For 
these reasons, results for the October run are not reported here. The December 
run included 65 $B$, $V$ image pairs, whereas the February run contained 
83 $B$, $V$ image pairs.

   \begin{figure*}
   \centering
   \includegraphics[width=12.5cm]{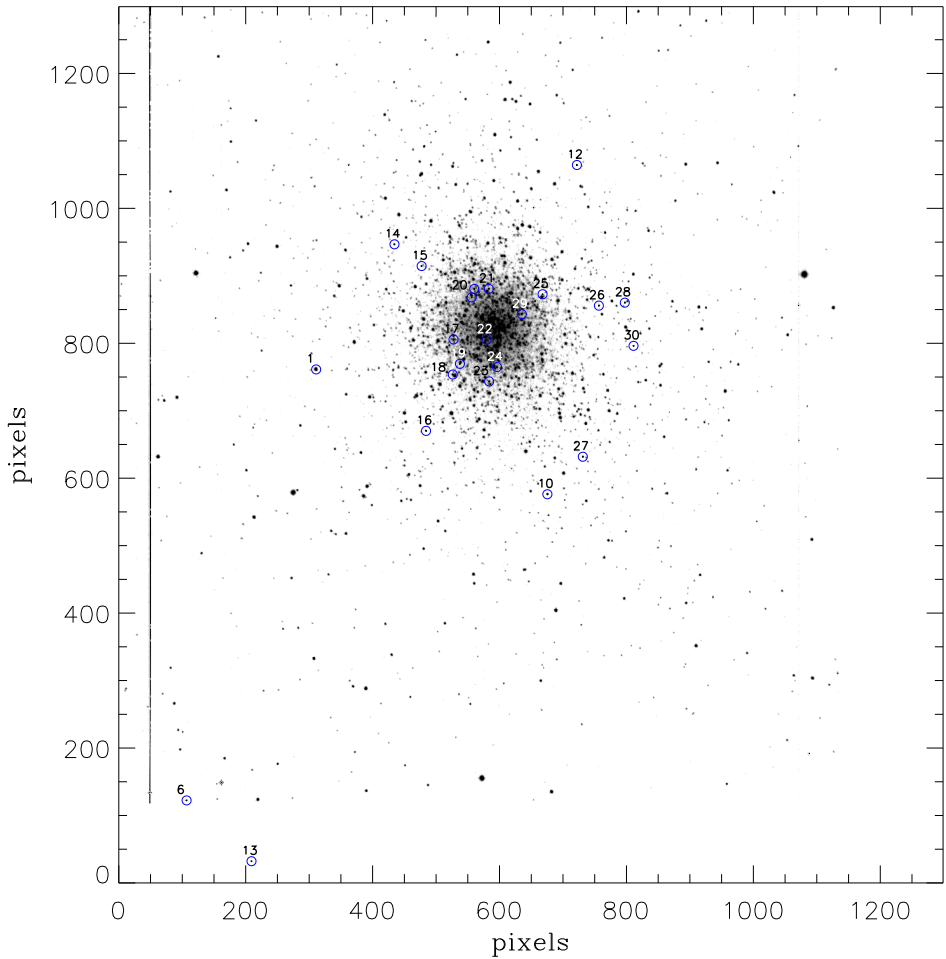}
      \caption{Finding chart for the NGC~2808 variables in our observed field. 
              }
         \label{Fig1}
   \end{figure*}

The ISIS2.1 analysis measures the difference in flux for stars in each 
image of the time series relative to their flux in a reference image 
obtained by stacking a suitable subset of images, and after convolving 
the images with a kernel to account for seeing variations and geometrical 
distortions of the individual frames (Alard 2000). We used the 10 $B$ 
images with the best seeing in each of the December and the February runs, 
respectively, to build up the reference images. In the case of $V$, we 
used 3 images obtained in the December run, closely spaced in time.  
Unfortunately, because of rotation 
between the two sets of images, we were not able to use the same 
reference image in $B$ for both of the runs. This resulted in offsets in 
the differential flux light curves for the two runs, which prevented us 
from consistently fitting $B$ light curves using the same periods as for 
the $V$ light curves, for which we were able to properly align the images
and use a single reference frame for both the December and February data. 
For these reasons, in what follows, unless otherwise noted, we will be 
addressing only the results based on the $V$ data. A full analysis of 
the $B$ data will be reported on in the future, along with a discussion 
of possible long-period variables.

   \begin{figure*}
   \centering
   \includegraphics[width=17cm]{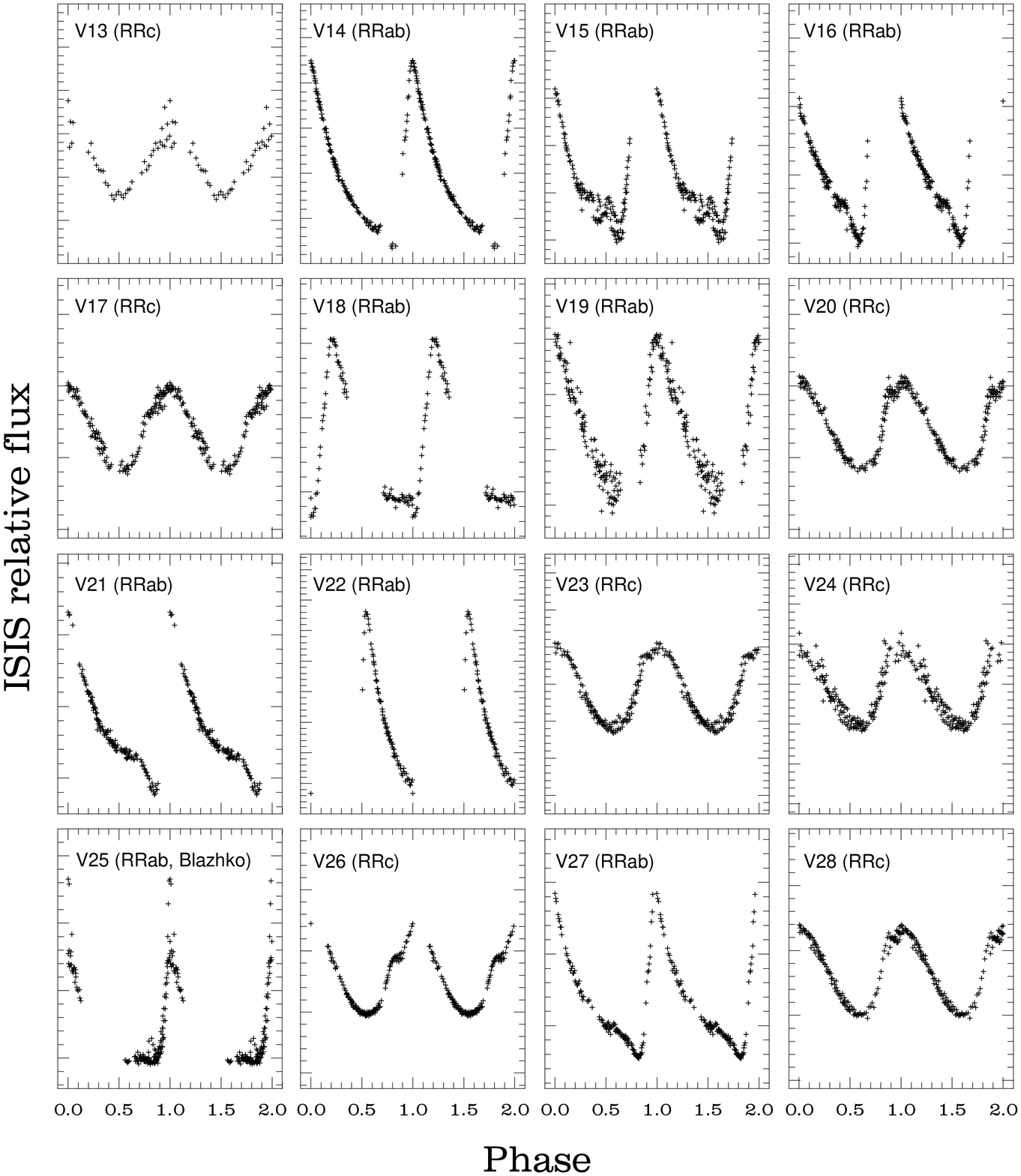}
      \caption{Light curves for newly discovered NGC~2808 RRL 
               variables. ISIS relative fluxes obtained using $V$-band 
               images were used in all cases except V18, V22, and V25.  
              }
         \label{Fig2}
   \end{figure*}

\section{Results} 
The ISIS analysis identified 18 new variables in NGC~2808, including 
16 RRL variables and three longer-period variables. 
Positional 
information and periods for these variables are provided in Table~1. 
Using the $V$-band data for the December and February runs, periods 
were computed using a variety of methods, including a period-finding 
program based on Lafler \& Kinman's (1965) ``theta'' statistic; the 
phase dispersion minimization technique (Stellingwerf 1978); information 
entropy minimization (Cincotta et al. 1995); Period98 (Sperl 1998); and 
CLEANest (Foster 1995). However, 
because of the relatively poor phase coverage and the long time interval
between the December and February observing runs, we have found that 
we are typically unable to decide among period aliases to within the 
$\pm 0.01$~d level, particularly in the cases of the RRab and longer-period 
stars. Therefore, periods in Table~1 are only provided to within this level 
of accuracy. 

Light curves for the newly detected RRL variables are provided in Fig.~2, 
where ISIS relative $V$ fluxes were used in all cases except when otherwise 
noted in the Appendix. Likewise, light curves for the longer-period variables, 
one of which (V30) is likely to be an eclipsing binary of the $\beta$ Lyrae 
type, are given in Fig.~3. The light curves for V1, V6, V10 and V12, the only 
previously known variable stars in NGC~2808 that were detected in our study, 
are shown in Fig.~4. Additional comments about individual variables can 
be found in the Appendix. The location and period of all the newly detected 
variables are given in Table~1, where the $x$ and $y$ 
values have the same meaning as in the Clement et al. (2001) catalogue.

\section{Discussion}
With our newly detected variables, we can recompute the specific 
frequency of RRL variable stars in NGC~2808. This quantity is defined as 

\begin{equation}
S_{\rm RR} = N_{\rm RR} \times 10^{0.4\, (7.5+M_V)},  
\end{equation}

\noindent where $M_V$ is the integrated visual magnitude of the cluster, 
or $M_V = -9.39$ according to the Harris (1996, Feb.~2003 update) catalogue. 
Given our newly discovered variables, one finds that the specific frequency 
of RRL variables in NGC~2808 is $S_{\rm RR} = 3.2$ (or 3.0, if V13 is not an 
RRL), compared with the value 0.3 as currently listed in the Harris catalogue. 

Even though the number of known RRL variables in the cluster has been 
increased by a factor of 10 with respect to previous estimates, the 
HB of NGC~2808 is still firmly classified as bimodal, since 
the numbers of both blue and red HB stars are much larger 
than the number of RRL present in the cluster (see, e.g., Table~1 in Lee 
et al. 1994). Likewise, the increase in the number of RRL stars implies 
but a small increase in the Lee-Zinn parameter that describes HB 
morphology in terms of the numbers of blue, red, and variable (RRL) 
stars; using the Lee et al. number counts as a guide, we estimate that 
the value provided in the Harris catalogue should be increased by only 
about 0.03.

   \begin{figure}
   \centering
   \includegraphics[width=5.5cm]{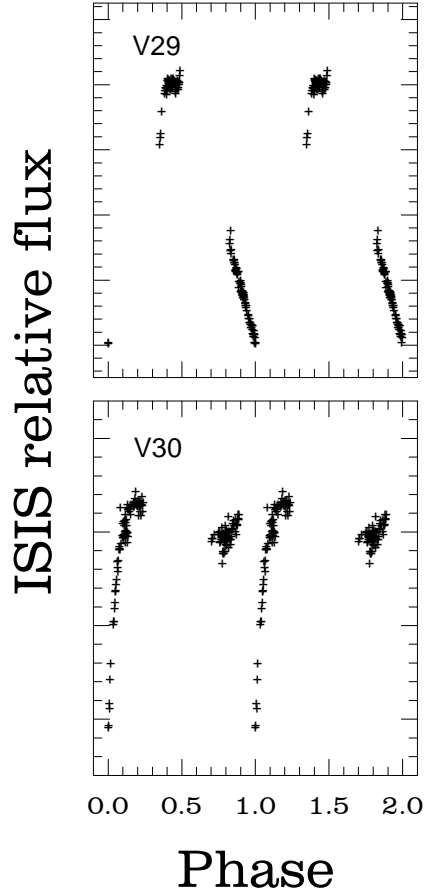}
      \caption{Light curves for newly discovered longer-period variables 
               in NGC~2808. ISIS relative fluxes obtained using $V$-band 
               images were used in both cases. V30 is likely to be a
               $\beta$ Lyrae-type EB.  
              }
         \label{Fig3}
   \end{figure}

As explained in the Introduction, future analysis of 
the detailed properties of the calibrated RRL light curves may reveal 
precious information about the origin of the bimodal HB of the cluster. 
In the future, we will attempt 
to employ resolved images of the core of the cluster, obtained from the 
ground or from space, to put the ISIS relative fluxes in the standard system, 
thus yielding reliable amplitudes and Fourier decomposition parameters for 
the cluster variables. The reason why this is not possible for most of the 
detected variable stars based solely on the available data is that we cannot 
obtain reliable photometry for these stars on our ISIS reference images, 
due to the extreme crowding that characterizes this cluster~--~one of the 
brightest in our galaxy. The difficulty in performing accurate photometry
in the inner cluster regions is, in fact, undoubtedly the main reason why a 
mere two RRL variables had previously been found towards NGC~2808 until we 
applied the image subtraction technique to our images. Note, in this 
sense, that even the finding chart for the cluster that we present 
in this Note (Fig.~1) is based on a 0.2~s exposure, obtained with the 
sole purpose of producing a finding chart for the crowded cluster fields 
where the variables were detected. 

In any case, we can already attempt a preliminary classification of the cluster 
into an Oosterhoff (1939) type on the basis solely of the mean period of its 
RRab and RRc stars, as well as on the c-to-ab number ratio.

The number fraction of c-type RRL, including the candidate RRe, 
is $N_{\rm c}/N_{\rm RR} \simeq 0.44$ (0.41, if V13 is not an RRL). 
This relatively large number 
ratio of c-type stars would clearly favor an Oosterhoff type II (OoII) 
classification for the 
cluster. On the other hand, the mean period of the 10 ab-type pulsators is 
$\langle P_{\rm ab}\rangle \simeq 0.563$~d, which indicates instead an 
OoI classification (see, e.g., Fig.~1a in Catelan 2004). For the 8 RRc and candidate RRe, 
the mean period is $\langle P_{\rm c}\rangle \simeq 0.30$~d (0.31~d, if V13 is not 
an RRL), which is also closer 
to the OoI type (e.g., Clement et al. 2001). Therefore, NGC~2808 presents characteristics 
of both Oosterhoff groups. As noted by Catelan (2004), however, more data are 
needed, particularly pulsation amplitudes, before the final word can be said in 
regard to the Oosterhoff type of the cluster.

\begin{acknowledgements}
We thank the referee for some valuable remarks, particularly in regard 
to V13's period, and Jos\'e Miguel Fern\'andez for his assistance during 
some of the observing runs. 
Support for M.C. was provided by Proyecto FONDECYT Regular No.~1030954.  
J.B. is supported by FONDAP Center for Astrophysics grant number 15010003. 
H.A.S. thanks the U.S. National Science Foundation for support under grant 
AST-0205813. 
Some of the analysis was performed with the computer program TS, developed 
by the American Association of Variable Star Observers. 
\end{acknowledgements}

   \begin{figure}
   \centering
   \includegraphics[width=10cm]{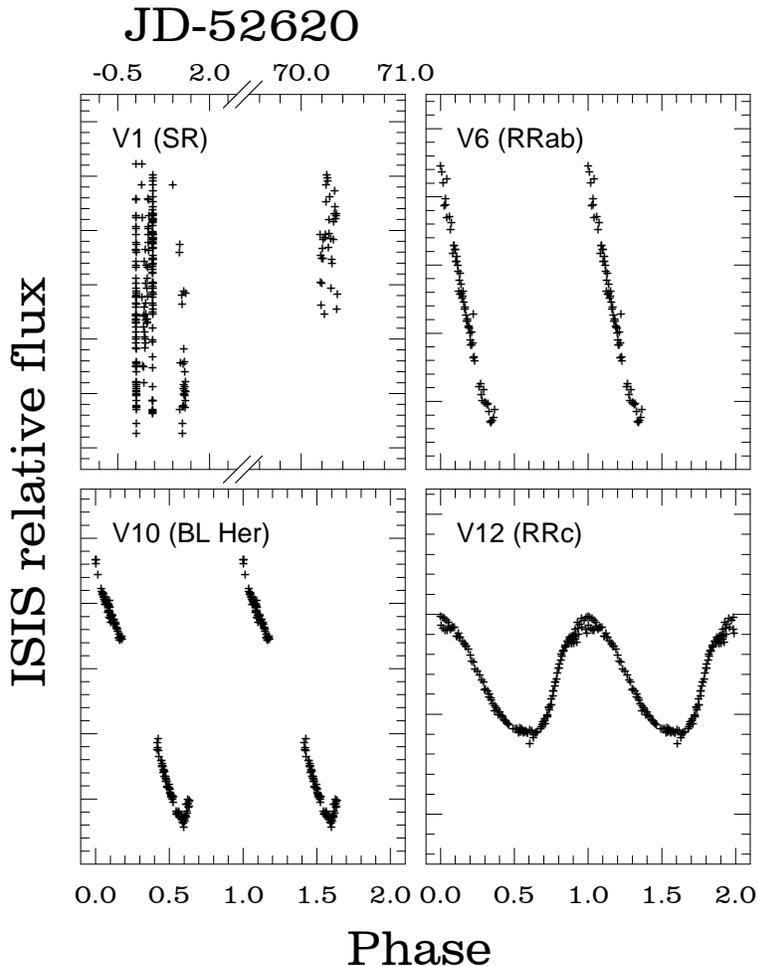}
      \caption{Light curves for V1 (upper left), V6 (upper right), V10 (lower 
               left) and V12 (lower right), the previously known variable 
               stars in NGC~2808 for which light curves could be obtained 
               in the present study. For V1, relative fluxes are plotted 
               as a function of the Julian Date (note the break in the 
               horizontal axis scale), whereas for other stars
               the light curves are phased using the period values indicated 
               in the main text. 
              }
         \label{Fig4}
   \end{figure}

\appendix 

\section{Notes on individual stars}

\subsection{V13} 
This star is located on the very southwestern edge of our field (Fig.~1). 
For this reason, it was not measured in about half of our images. The period 
that best phases the current data is very short; therefore, if not an RRc, 
this star might be an RRe~--~i.e., a second-overtone pulsator. One less  
likely possibility, which however cannot be conclusively ruled out by our 
data due to our poor phase coverage, is that this star is actually a 
W Ursae Majoris eclipsing binary (EB), with a period perhaps around 0.44~d. 

\subsection{V18} 
A good light curve for this star could not be obtained in $V$. In Fig.~2 is 
shown only the light curve obtained on the basis of the February $B$ data. 

\subsection{V19}
The $V$ light curve presents significantly more scatter than the $B$ light 
curve for each individual month, which could be due to imperfect subtraction 
of nearby bright RGB stars. 

\subsection{V22}
Again like V19, the $V$ light curve presents significantly more scatter than 
the $B$ light curve for each month, which could also be due to imperfect 
subtraction of nearby bright red giants. In fact, this variable is the one 
closest to the cluster centre to be detected in the present study (Fig.~1). 
In Fig.~2 is shown only the light curve obtained on the basis of the February 
$B$ data.

\subsection{V25}
Both the December and February $B$ data are reasonably phased with a period 
near 0.49~d and an ab-type shape. However, one finds a large difference in 
amplitude between the two months. In addition, the two maxima appear slightly 
dephased. The star may be a Blazhko variable. To illustrate this, 
we show the star's light curve in $B$ in Fig.~2, where both the February 
and December data were employed, but a single period was used. 
The $V$ light curve presents significantly more scatter than the $B$ light 
curve for each individual month, which could be due to imperfect subtraction 
of nearby bright RGB stars. 

\subsection{V29}
This appears to be a longer-period variable, possibly a Cepheid. Our phase 
coverage is insufficient to derive a reliable period and light curve for this 
star. A light curve for a representative candidate period is  
shown in Fig.~3 (top panel). 

\subsection{V30}
The light curve resembles that of an EB, most likely
of the $\beta$~Lyrae type. A W Ursae Majoris classification is also possible. 
However, our phase coverage seems to have been insufficient to derive a reliable 
period and light curve. A light curve for our best candidate period (Table~1) 
is shown in Fig.~3 (middle panel). 

\subsection{Previously known variables}
Of the previously known variables in the cluster, we have only been able to 
obtain a complete light curve for the RRc variable V12, which is shown in Fig.~4 
(right panel) using the same period as in Clement \& Hazen (1989), or 0.30577705~d. 
This period clearly phases our data 
adequately. Clement et al. (2001) currently classify V12 as a possible RRe; 
we are unable at present to decide whether this is more suitable than a more 
straightforward RRc-type classification, and have thus adopted the latter until 
more data are available, including Fourier parameters and $I$-band magnitudes 
(Catelan 2004 and references therein). V1, which is also in our field, is 
classified as a semi-regular (SR) variable. In Fig.~4, relative fluxes for this 
star are plotted as a function of the Julian Date. Obviously, our 
phase coverage is necessarily incomplete for this star. V6 is an RRab lying at 
the southwestern edge of our image; as it happens, it only fell within our field 
of view in the February run, resulting in an incomplete light curve as can also 
be seen in Fig.~4~--~where we have adopted the Clement \& Hazen period of 
0.5389687~d, which seems adequate. V10, for which we again were able to obtain 
only an incomplete light curve, is classified as a BL Her variable with a period 
1.76528~d. This period appears to describe our data satisfactorily (Fig.~4).

\end{document}